\documentclass[preprint,12pt]{elsarticle}

\usepackage{amssymb}
\usepackage{lineno}
\usepackage{siunitx}
\usepackage{hyperref}
\hypersetup{colorlinks,allcolors=black} % for arXiv

\usepackage{wasysym} % \diameter

\journal{Nuclear Instruments and Methods in Physics Research Section A}

\begin{document}

\begin{frontmatter}



\title{Development of a linac-based LEPD experimental station for surface structure analysis and coordination with synchrotron radiation ARPES}


\author[inst1]{Rezwan Ahmed}
\author[inst1]{Izumi Mochizuki}
\author[inst1]{Toshio Hyodo}
\author[inst2]{Tetsuroh Shirasawa}
\author[inst1]{Seigi Mizuno}
\author[inst3]{Yoshinari Kondo}
\author[inst1]{Kenichi Ozawa}
\author[inst1,inst4]{Miho Kitamura}
\author[inst1]{Kenta Amemiya}
\author[inst5]{Bartlomiej Checinski}
\author[inst5]{Jozef Ociepa}
\author[inst6,inst7]{Achim Czasch}
\author[inst6,inst7]{Ottmar Jagutzki}
\author[inst1]{Ken Wada\corref{cor1}}

\affiliation[inst1]{organization={Institute of Materials Structure Science, High Energy Accelerator Research Organization (KEK)},%Department and Organization
            addressline={1-1 Oho}, 
            city={Tsukuba},
            postcode={Ibaraki 305-0801}, 
            %state={},
            country={Japan}}
\affiliation[inst2]{organization={Research Institute for Material and Chemical Measurement, National Institute of Advanced Industrial Science and Technology (AIST)},%Department and Organization
            addressline={1-1-1 Higashi}, 
            city={Tsukuba},
            postcode={Ibaraki 305-8565}, 
            %state={},
            country={Japan}}
\affiliation[inst3]{organization={Support Center for Accelerator Science and Technology, High Energy Accelerator Research Organization (KEK)},%Department and Organization
            addressline={1-1 Oho}, 
            city={Tsukuba},
            postcode={Ibaraki 305-0801}, 
            %state={},
            country={Japan}}
\affiliation[inst4]{organization={NanoTerasu Center, National Institutes for Quantum Science and Technology (QST)},%Department and Organization
            addressline={6-6-11 Aoba}, 
            city={Sendai},
            postcode={Miyagi 980-8579}, 
            %state={},
            country={Japan}}
\affiliation[inst5]{organization={OCI Vacuum Microengineering Inc.},%Department and Organization
            addressline={200 Stronach Crescent}, 
            city={London},
            postcode={N5V 3A1}, 
            state={Ontario},
            country={Canada}}
\affiliation[inst6]{organization={Institut für Kernphysik, Universität Frankfurt},%Department and Organization
            addressline={Max-von-Laue-Str. 1}, 
            city={Frankfurt},
            postcode={D-60438}, 
            state={Hesse},
            country={Germany}}
\affiliation[inst7]{organization={RoentDek Handels GmbH},%Department and Organization
            addressline={Im Vogelshaag 8}, 
            city={Kelkheim},
            postcode={D-65779}, 
            state={Hesse},
            country={Germany}}

\cortext[cor1]{kenwada@post.kek.jp}

\begin{abstract}
We report on the development of a low-energy positron diffraction (LEPD) experimental station for surface structure analysis using a linear electron accelerator (linac)-based slow-positron beam. LEPD is the positron counterpart of low-energy electron diffraction (LEED) and is expected to offer higher accuracy in surface structure determination compared to LEED. The newly developed station enables acquisition of LEPD $I$-$V$ curves (where the intensity of each inequivalent diffraction spot is shown as a function of the incident beam energy) for a sample within a few hours, thus making measurements feasible before surface degradation occurs. The station consists of two ultra-high vacuum (UHV) chambers, separated by a gate valve: one for sample preparation and the other for LEPD $I$-$V$ measurements. The preparation chamber is equipped with an Ar$^+$ sputtering system for sample cleaning, a triple-pocket electron beam evaporator for adsorbate deposition, three gas introduction systems, additional ports available for user-specific needs, as well as a LEED/Auger electron spectroscopy (AES) system. The LEPD chamber, equipped with LEPD optics and a detector, is connected to the linac-based slow-positron beamline. Sample manipulators installed in both of the chambers are designed to enable rapid sample cooling, as well as precise positioning and orientation adjustments. In the preparation chamber, the manipulator is additionally capable of direct current heating of a sample up to \qty{1200}{\degreeCelsius}. The sample holder is compatible with this LEPD station at SPF-A4 of the Slow Positron Facility (SPF) and the angle-resolved photoemission spectroscopy (ARPES) station at BL-13B of the Photon Factory (PF), both located at the Tsukuba campus of the Institute of Materials Structure Science (IMSS), High Energy Accelerator Research Organization (KEK). The design concepts are described along with experimental demonstrations. 
\end{abstract}




\end{frontmatter}

%\linenumbers

\section{Introduction}

Over the past several decades, low-energy electron diffraction (LEED) has been widely used to determine surface atomic arrangements. In LEED, a low-energy electron beam, typically with energies ranging from tens to hundreds of eV, is projected onto the sample surface typically in the direction perpendicular to the surface, and the resulting back-scattered diffraction pattern is observed. While LEED is often employed for quick checks of the two-dimensional periodicity of the surface lattice, it is also used for the precise determination of surface atomic arrangements through quantitative $I$-$V$ analysis. This involves measuring the intensity of each inequivalent diffraction spot as a function of the incident beam energy and comparing with theoretical calculations~\cite{pendry1974low,vanhove1986low}. Although it is a well-established method, limitations in the accuracy of multiple-scattering LEED calculations can result in unsatisfactory structure determination.

Low-energy positron diffraction (LEPD), which uses the antiparticle of the electron, is considered to have several characteristics that offer advantages over LEED for surface structure analysis. These characteristics include a simple and smooth atomic scattering factor, reduced multiple scattering, and a shorter inelastic mean free path, which leads to higher surface sensitivity~\cite{tong2000positron}. Nearly half a century ago, Canter et al.\ at Brandeis University reported the first LEPD observation using an angled incidence positron beam generated from a radioisotope (RI) source and positron detection with a channel electron multiplier (CEM)~\cite{rosenberg1980low}. Around the same time, Mills and Platzman at Bell Laboratories independently reported their LEPD observation~\cite{mills1980observation}. Subsequent studies were conducted by research groups at University of Texas at Arlington and Brookhaven National Laboratory, alongside continued efforts at Brandeis University and Bell Laboratories, exploring LEPD with various configurations and detection methods~\cite{jona1980theory,coleman1982observation,weiss1983low,cook1984elastic,frieze1985positron,mills1985low,mayer1987low}. Later, Canter and collaborators further advanced the subject by developing a normal-incidence LEPD system capable of multi-spot observation to further enhance surface structure analysis capabilities~\cite{horsky1989observation}. Despite the limitation in beam intensity from an RI-based source, their pioneering work successfully demonstrated better agreement between experimental and calculated $I$-$V$ curves than LEED~\cite{horsky1989observation,duke1989surface,horsky1992analysis,lessor1991low,chen1993low,duke1997low}. 

A high-intensity pulsed slow-positron beam is provided at the Slow Positron Facility (SPF) in the Institute of Materials Structure Science (IMSS) at the High Energy Accelerator Research Organization (KEK), using a dedicated linear electron accelerator (linac)~\cite{wada2012increase,wada2013new,hyodo2018slow}. Experiments on total-reflection high-energy positron diffraction (TRHEPD), the positron counterpart of reflection high-energy electron diffraction (RHEED), which preceded the development of the LEPD station discussed in the present study, were made possible by this high-intensity positron beam. The TRHEPD station~\cite{maekawa2014brightness}, located at beamline SPF-A3, has been in operation for SPF users for over a decade, delivering a number of significant results and demonstrating high accuracy in elucidating the atomic geometry of the outermost and underlying atomic layers~\cite{fukaya2019totala,endo2020structure,fukaya2021atomic,tsujikawa2022structural,tsujikawa2022structure,fukaya2023reversible,dodenhoeft2023determination}. The LEPD station was later developed at SPF-A4~\cite{wada2018observation,wada2020pulse}. The higher-intensity linac-based slow-positron beam, compared to the RI-based one, is expected to enhance this capability for surface atomic structure determination. 

We have improved on the previous LEPD station, enhancing its capabilities to facilitate practical LEPD $I$-$V$ measurements for surface structure analysis. Additionally, a new type of LEPD sample holder has been developed, which is compatible with the angle-resolved photoemission spectroscopy (ARPES) station at beamline BL-13B~\cite{ozawa2022beamline} of the Photon Factory (PF) in IMSS at KEK. This holder allows both LEPD and ARPES experiments to be performed on the same sample. ARPES enables the determination of the electron energy dispersion relation, i.e., the band structure, of crystalline materials. The combined use of LEPD and ARPES allows for a comprehensive determination of both atomic and electronic structures on the surface, providing deeper insights into the physical and chemical properties of surfaces.

\section{The LEPD system using a linac-based slow-positron beam at the SPF, IMSS, KEK}

\begin{figure}[thbp]
    \centering
    \includegraphics[width=1.0\linewidth]{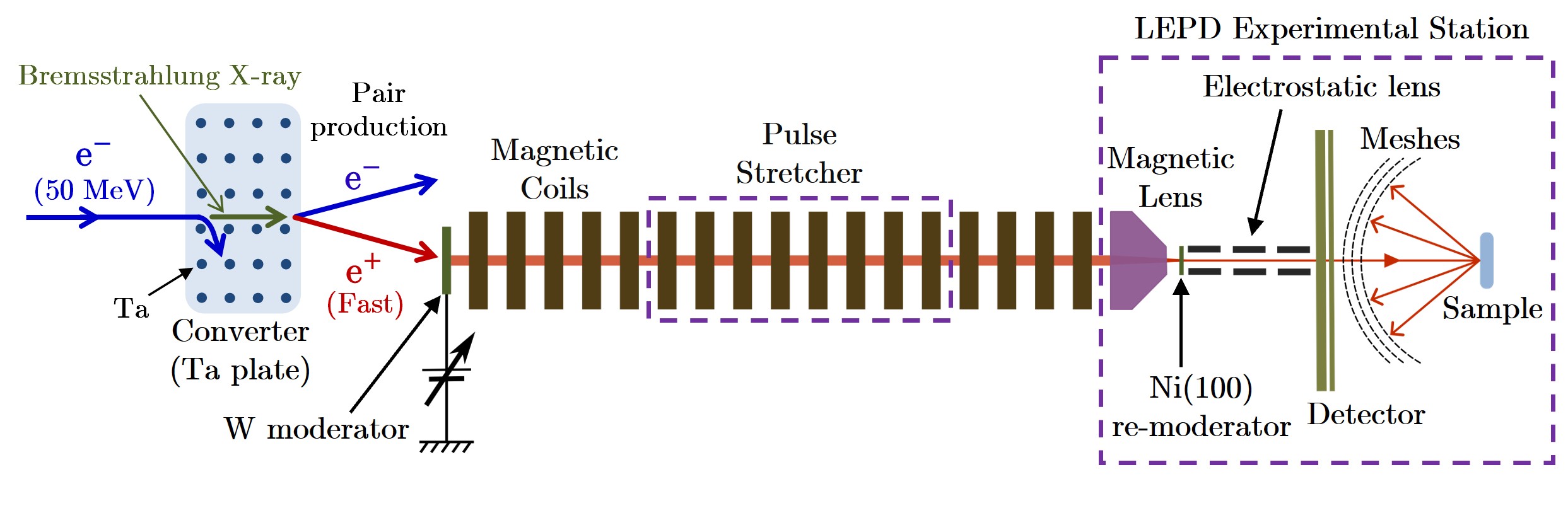}
    \caption{Schematic diagram of the low-energy positron diffraction (LEPD) experiment system with a linac-based slow-positron beam at the Slow Positron Facility, IMSS, KEK.}
    \label{fig:systemdiagram}
\end{figure}

The conceptual diagram of the LEPD system using the linac-based slow-positron beam at the SPF, IMSS, KEK~\cite{wada2018observation,wada2020pulse} is shown in Fig.~\ref{fig:systemdiagram} and briefly described here. The high-intensity pulsed slow-positron beam is generated using a dedicated linac operated at \qty{50}{MeV}, \qty{530}{W}, with a repetition frequency of \qty{50}{Hz}~\cite{wada2012increase}. The accelerated electrons are injected into a tantalum (Ta) converter, where they are deflected by Ta nuclei, emitting Bremsstrahlung X-rays. A portion of this radiation, with energy exceeding a threshold of twice the rest-mass energy of the electron, produces positron-electron pairs in the same Ta converter, initiating a cascade shower of positron-electron pair production. A fraction of the resulting high-energy positrons, with energies ranging up to ${\sim}\qty{50}{MeV}$, thermalizes in tungsten (W) foils with a thickness of \qty{25}{\micro\meter}, and is re-emitted from the surfaces with an energy of ${\sim}\qty{3}{eV}$, corresponding to the negative positron work function of W. These mono-energetic re-emitted positrons are referred to as slow positrons~\cite{mills1983experimentation,schultz1988interaction}.

The slow positrons are immediately accelerated to \qty{5}{keV} and transported along a magnetic field generated by coils aligned with the beamline duct, which extends into the experimental hall. Midway through the beamline, the pulse width is stretched~\cite{wada2020pulse} to match the LEPD detector’s limited simultaneous multi-detection capability. The pulse stretcher adjusts the pulse width over a range of approximately \qty{200}{\micro\s} to \qty{20}{ms}, with the latter corresponding to an almost continuous beam. 

The stretched beam, delivered at a slightly increased energy of \qty{5.2}{keV}, is transported to the LEPD station and focused onto a Ni(100) film using a magnetic lens outside the beam transport coils, where it is re-moderated to enhance the beam brightness~\cite{mills1982transmission,zafar1989single,fujinami2008production,wada2018observation}. The re-moderated positrons, having a kinetic energy \qty{1}{eV} higher than the potential applied to the re-moderator due to the negative work function of the Ni film~\cite{zafar1989single}, are subsequently transported by an electrostatic lens system and directed through a thin metal tube to the sample. The tube passes through the central hole of a position-sensitive detector equipped with a microchannel plate (MCP) assembly and retarding meshes, before the positrons normally impinge on the surface of the crystalline sample. The diffracted positrons are then detected by the position-sensitive detector, where both their position coordinates and relative timing information are encoded.

\section{Upgraded LEPD experimental station}

The LEPD diffraction patterns of the Ge(001)--$2\!\times\!1$ structure were previously observed at the SPF, though several hours were required to accumulate sufficient data for each pattern \cite{wada2018observation}. The experimental station at that time faced limitations in both sample preparation and the precise adjustment of sample orientation and cooling, which are critical for quantitative $I$-$V$ analysis. The prolonged adjustments, in addition to the waiting time for cooling, resulted in surface contamination by residual gases, weakening the diffraction spot intensities and hindering accurate surface structure determination. 

\begin{figure}[htbp]
    \centering
    \includegraphics[width=1.0\linewidth]{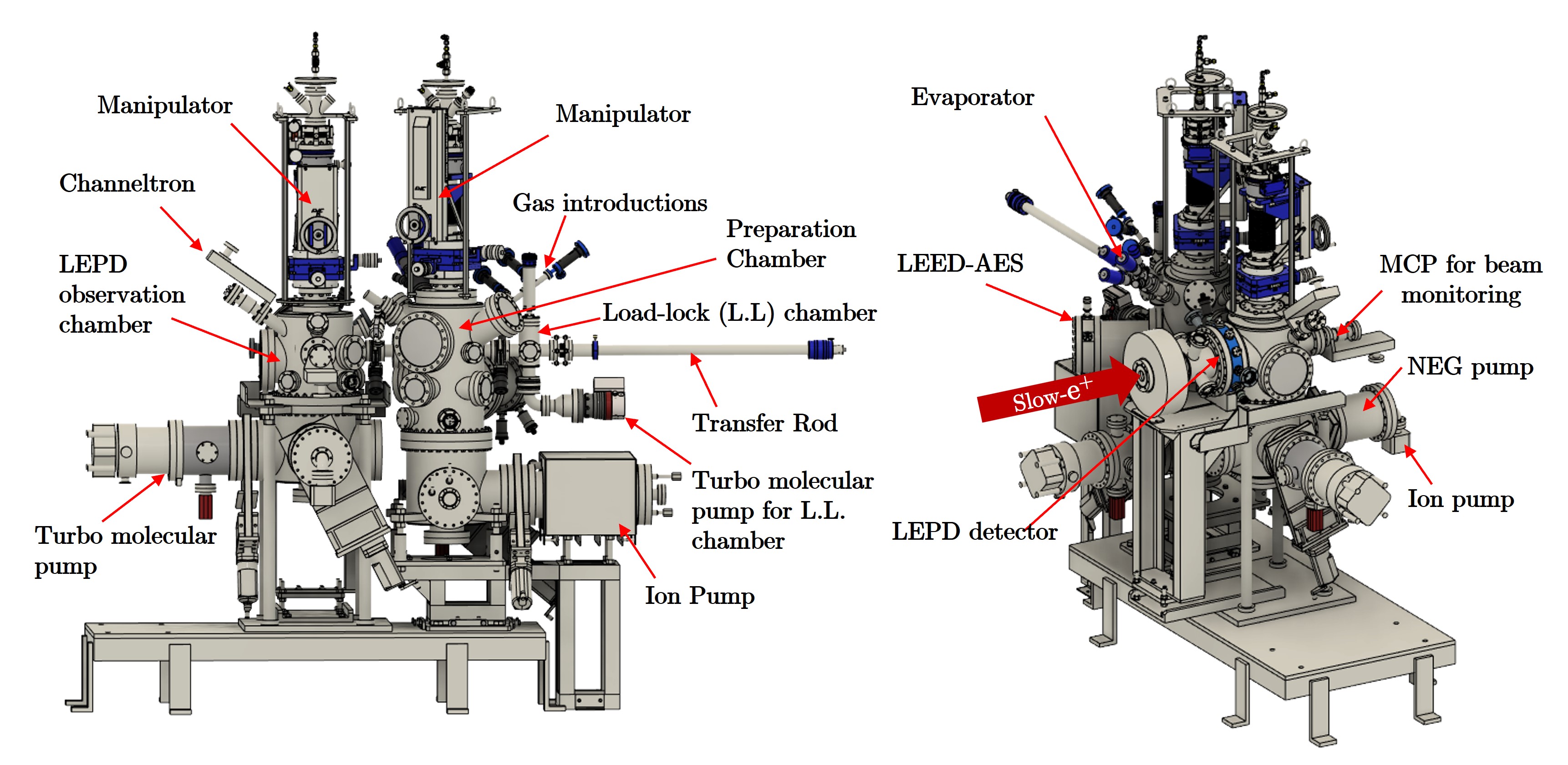}
    \caption{Front (left) and corner (right) views of the upgraded LEPD experimental station connected to the SPF-A4 beamline. It consists of two UHV chambers. Some of the components connected to the ports of the two chambers are indicated.}
    \label{fig:lepdstation}
\end{figure}

The upgraded LEPD experimental station addresses these issues by incorporating a more efficient sample preparation equipment and systems that reduce contamination risks. Fig.~\ref{fig:lepdstation} shows the upgraded experimental station, which consists of a ``sample preparation chamber" and a ``LEPD observation chamber", the latter connected to the SPF-A4 beamline branch. The design ensures seamless integration between sample preparation and LEPD measurements.

\subsection{Sample preparation chamber}

The sample preparation chamber is designed for efficient surface preparation and is equipped with LEED and Auger electron spectroscopy (AES) optics for surface condition analysis. A load-lock system, connected to this chamber by a gate valve, allows for the pre-loading of up to five sample holders in a sample stocker, enabling speedy sample exchange. Once the load-lock chamber has been pumped for approximately six hours using a dedicated turbomolecular pump (TMP) coupled with a dry pump after initial sample loading from the atmosphere, subsequent sample exchanges can be performed quickly without requiring additional pumping. Samples are transferred from the load-lock chamber to the sample preparation chamber using a transfer rod with newly developed reliable catch-and-release mechanisms. The load-lock system, combined with this sample transfer rod, ensures efficiency and consistency in sample preparation. Without this system, if a sample preparation failure occurred and a sample exchange were needed, several days of baking of the large chamber would be required for the new attempt. This could result in the loss of valuable beam time, which is available only a limited number of times for each user project per year. 

The chamber is also equipped with an Ar$^+$ sputtering system for sample cleaning, a triple-pocket electron beam evaporator for atom deposition, three gas introduction systems, and additional ports available for user-specific needs. The UHV conditions of the preparation chamber are maintained by an ion pump and a TMP coupled with a dry pump, both of which can be isolated using gate valves when necessary. 

\subsection{LEPD measurement chamber and Hexanode detector upgrade}

The LEPD chamber, made from SUS316L stainless steel, is surrounded by correction coils to cancel the Earth's magnetic field. Its inner surface was finished with chemical-mechanical polishing, and the entire chamber was dehydrogenated by annealing at \qty{450}{\degreeCelsius} for \qty{48}{hours}. After assembly, the chamber was initially evacuated using a TMP and dry pump set, followed by baking to achieve UHV conditions. During the cooling phase of the baking, a non-evaporative getter (NEG) pump was activated through in-situ heating. The NEG pump can be isolated from the chamber by a gate valve when necessary, such as during chamber exposure to atmosphere, eliminating the need for reactivation after initial activation. The UHV of the chamber is primarily maintained by the NEG pump. Slowly emerging noble gases that the NEG pump cannot absorb are captured by a small ion pump positioned in the rear of the NEG pump and separated by a partition with a \qty{3}{mm} diameter hole. The conductance of the hole is small enough to minimize detector background due to the charged particles leaking from the ion pump, while being sufficient to prevent the accumulation of noble gases in the chamber. The TMP and dry pump set is isolated by a gate valve whenever the positron beam is fed into the chamber, safeguarding the vacuum of the entire beamline, including the linac, in case of a sudden power failure. 

One of the major limitations in the previous LEPD system was the large cross-shaped dead zone in the detector~\cite{wada2018observation}, where no signals, including diffraction spots, could be observed. The delay-line anode detector used in the previous setup had two anode wires wrapped perpendicularly to each other around a support placed behind the MCP. The position was determined from the time difference between signals, caused by electron clouds amplified by the MCP, reaching both ends of the wires. Since there was a hole at the center of the detector for the beam to pass through, the anode wires had to avoid this, resulting in gaps in both the vertical and horizontal layers, which formed a cross-shaped dead zone. To determine both $x$ and $y$ coordinates, each point on the detector must be covered by two layers of wires. Recovering the lost spots with this configuration would require additional mechanisms for sample rotation, as well as extended adjustments and measurement times. 

This issue has been resolved by incorporating a three-layer delay-line anode detector, referred to as a ``Hexanode" after its hexagonal appearance~\cite{jagutzki2002multiple}. In the Hexanode, three sets of anode wires are wrapped at \ang{120} angles to each other, ensuring that each point on the detector is covered by at least two layers of wires. This design eliminates the cross-shaped dead zone and enables more comprehensive LEPD pattern acquisition. The Hexanode has an effective image diameter of approximately \qty{75}{mm} with a central hole~\cite{roentdekweb}. Its maximum outer dimension is \qty{196}{mm}. It is enclosed by a $\mu$-metal cover in order to shield it from earth and stray magnetic fields. The outer diameter of the cover is \qty{204}{mm}, which provides a minimum distance from the high-voltage components sufficient to prevent discharge. (During operation, high voltages of up to approximately \qty{3}{kV} are applied at specific parts of the Hexanode.) However, this exceeds the inner diameter of a standard DN200CF (ICF253) flange port. Hence, a special port with an inner diameter of \qty{210}{mm} was employed to accommodate the Hexanode on the LEPD measurement chamber. The detector, along with the optics and retarding meshes, was mounted on the DN200CF flange with double-sided edges, ensuring proper alignment and wiring.

Initially, micro-discharges occurred, and due to the detector’s extreme sensitivity, many hot spots appeared in the detected image, as the detector counted every charged particle arriving at the sensitive area. These discharging problems were resolved by eliminating unnecessary insulating materials that contacted both ground and high-voltage parts. 

The LEPD chamber is also equipped with an MCP-phosphor screen assembly for beam monitoring and a channel electron multiplier (CEM) assembly for measuring the incident beam intensity by counting positrons. These detectors are mounted on two independent linear transfer stages, allowing them to be positioned at the same location as the sample by moving the sample aside. This setup enables separate checks of the beam shape and intensity prior to and/or after LEPD observations. 

\subsection{Manipulators and sample holders}

Each chamber is equipped with a compact manipulator. The manipulator in the LEPD observation chamber allows
horizontal (perpendicular to the manipulator rod) translations within $\pm \, \qty{12.5}{mm}$, with a precision of \qty{0.005}{mm}, and vertical (along the rod) translation of $\qty{100}{mm}$, with a precision of \qty{0.1}{mm}. The manipulator in the sample preparation chamber offers the same translation capability but with a vertical translation of $\qty{200}{mm}$, allowing it to cover both the sample preparation space in the upper part of the chamber and the LEED/AES observation space in the lower part of the chamber. Both manipulators rotate a full \ang{360} around the vertical axis, with an angular precision of \ang{0.05}. The sample tilt angle can be adjusted up to $\pm \, \ang{2}$, with a precision of \ang{0.078}.

The sample plate receiver, attached to the manipulator in the preparation chamber, is wired to the top feedthrough of the manipulator rod, allowing resistive heating with a current of up to \qty{18}{A}. The cables connected to the feedthrough rotate with the manipulator rod, enabling unrestricted rotation in either direction. Two contact electrodes of the sample plate receiver secure the electrodes of the sample holder by springs. The sample holders and receivers are designed to be compatible with those used in the surface science experimental stations at the Photon Factory (PF), IMSS, KEK. 

\begin{figure}[htbp]
    \centering
    \includegraphics[width=0.4\linewidth]{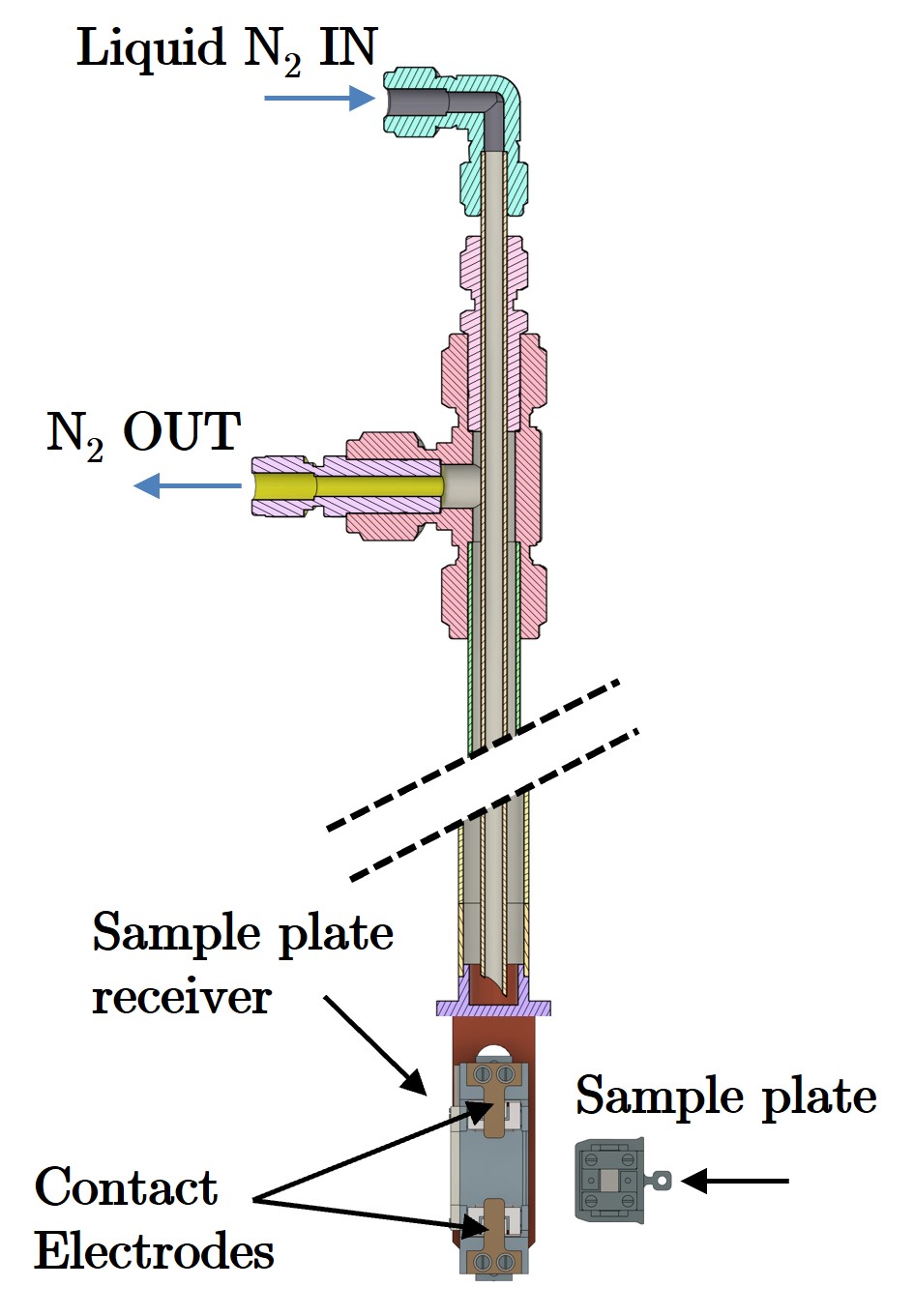}
    \caption{Schematic cross-sectional view of the manipulator rod and sample plate receiver}
    \label{fig:manipulater}
\end{figure}

During the LEED observations in the preparation chamber and the LEPD observations in the LEPD chamber, the sample is cooled to below \qty{-180}{\degreeCelsius}, a standard procedure for $I$-$V$ measurements that suppresses the effect of atomic thermal vibrations as described by the Debye-Waller factor, and enhances diffraction beam intensity. The compact design of the manipulators ensures rapid cooling, which is critical for minimizing surface contamination before $I$-$V$ measurements. The manipulator rod is hollow to accommodate liquid nitrogen (N$_2$), as illustrated in the cross-sectional diagram shown in Fig.~\ref{fig:manipulater}. During the experiment, liquid N$_2$ is continuously supplied from a liquid N$_2$ vessel to a stainless-steel tube inside the hollow rod, which makes direct contact at the bottom with an oxygen-free copper block, on which the sample plate receiver is mounted. The evaporated N$_2$ is vented through another opening at the top of the manipulator.

\begin{figure}[htbp]
    \centering
    \includegraphics[width=1.0\linewidth]{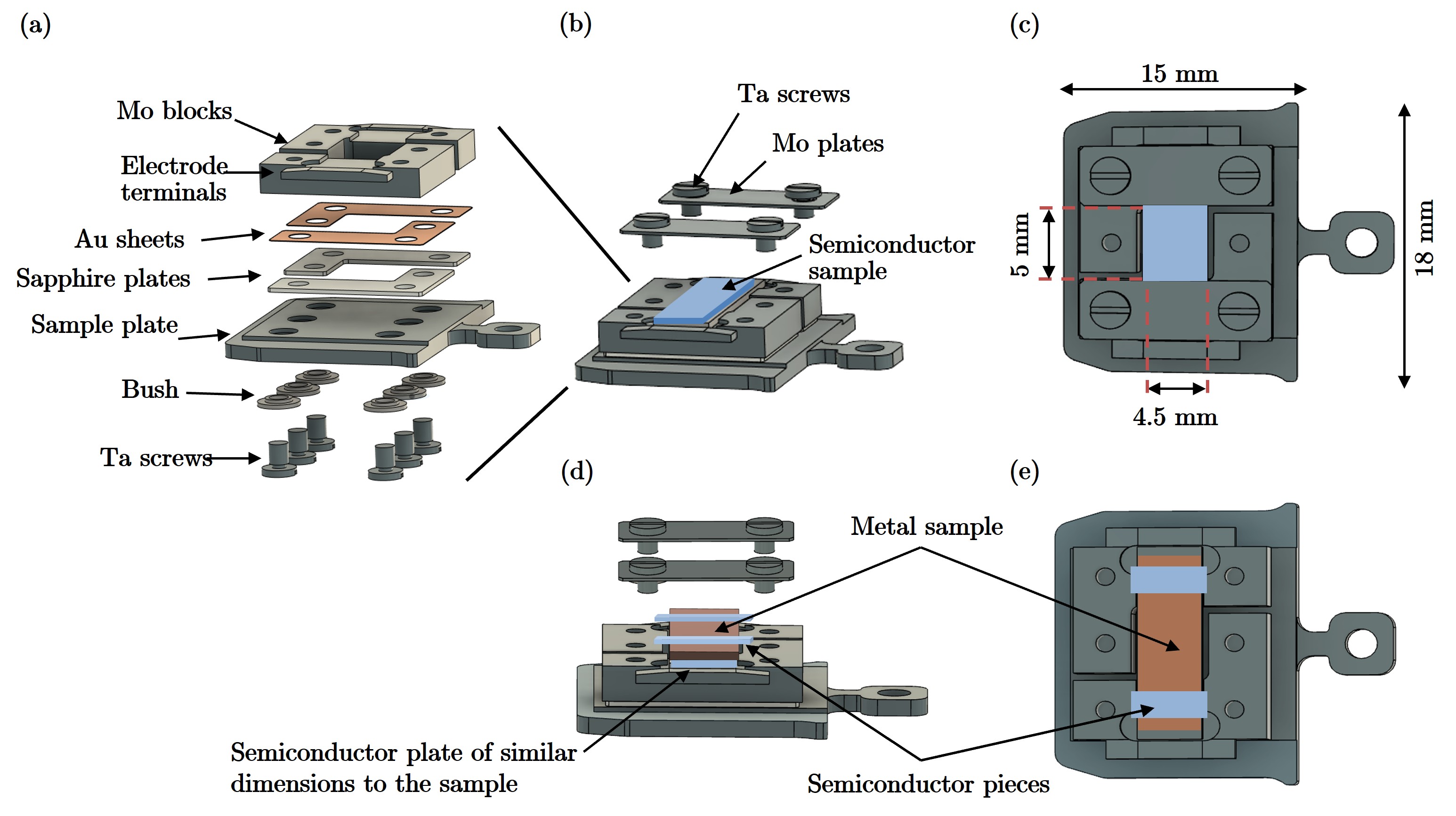}
    \caption{(a) An exploded view of the components of the sample holder, designed for both efficient cooling and heating. The holder consists of a molybdenum (Mo) sample plate, sapphire plates, gold (Au) sheets, and Mo blocks, which are secured to the sample plate from the bottom side using screws and insulating bushes. (b) A semiconductor sample is mounted on the Mo blocks and fixed by Mo plates and tantalum (Ta) screws. (c) Top view of the sample holder with a semiconductor sample mounted, showing the $\qty{4.5}{mm} \times \qty{5}{mm}$ central area ready to be exposed to the incident beam. (d) A metal sample is fixed by Mo plates and tantalum (Ta) screws, with a semiconductor plate of similar dimensions placed underneath the sample and small semiconductor pieces positioned on the top surface at both ends of the sample for resistive heating. (e) Top view of the sample holder before the Mo plates are fixed, showing a mounted metal sample and semiconductor pieces on the top surface at both ends. The maximum sample size accommodated by this sample holder is an area of $\qty{4.5}{mm}\times\qty{12}{mm}$ with a thickness of approximately \qty{1.5}{mm}.}
    \label{fig:sampleholder}
\end{figure}

Fig.~\ref{fig:sampleholder} shows a schematic diagram of the components of the sample holder. The sample is fixed on the molybdenum (Mo) blocks by two Mo plates with tantalum (Ta) screws, as shown in Fig.~\ref{fig:sampleholder}~(b). Screws of appropriate length are used in accordance with the thickness of the sample including any additional layers required for specific sample setups (e.g., heating of metallic samples, as described below), so that they do not touch the screws securing the Mo blocks to the sample plate from its bottom side. In this arrangement, a central area of $\qty{4.5}{mm} \times \qty{5}{mm}$ is open for the incident beam, as shown in Fig.~\ref{fig:sampleholder}~(c). The maximum dimensions of the sample that can be accommodated in this sample holder are an area of $\qty{4.5}{mm} \times \qty{12}{mm}$ and a thickness of approximately \qty{1.5}{mm}.

Two sapphire plates and gold (Au) sheets are placed between the sample plate and the two Mo blocks which are secured to the sample plate while insulated from it using screws and insulating bushes, as shown in Fig.~\ref{fig:sampleholder}~(a). The sapphire plates provide electrical insulation with a moderate heat conduction, while the Au sheets improve thermal contact. This electrical isolation enables the Mo blocks---and consequently the sample---to float from ground potential. 

In order to optimize the position and orientation of the sample for Ar$^+$ sputtering in the sample preparation chamber, the ion current is monitored via an ammeter connected in series between the Mo blocks and the grounded chamber through the feedthrough. The determined optimal position and orientation are recorded as the manipulator's position and angle scale values. Once this is done, future sputtering can be performed by reproducing the sample's position and orientation referring to the recorded values without further ion current monitoring. LEED/AES measurements in the sample preparation chamber are made with the sample grounded externally via the feedthrough. In the LEPD observation chamber, the sample is locally grounded through the Mo blocks and the manipulator to the chamber.

Semiconductor samples are resistively heated directly up to \qty{1200}{\degreeCelsius} by passing current between the Mo blocks along the sample. Metallic samples, on the other hand, are indirectly heated using the resistive heating of a semiconductor. In these instances, a semiconductor plate of similar dimensions to the metallic sample is placed underneath it, while small semiconductor pieces are positioned on top of the sample at both ends. These pieces are secured in place by the Mo top plates with screws of appropriate length, as shown in Fig.~\ref{fig:sampleholder}~(d) and Fig.~\ref{fig:sampleholder}~(e). The semiconductor plate beneath the sample also serves to prevent it from sagging when heated. The current flows from one Mo block through the part of the semiconductor plate in contact with the block, as well as through the top semiconductor pieces into the metal sample. It then flows along the sample, and returns through the opposite section of the semiconductor plate and the piece on the opposite side into the other Mo block. This flow generates resistive heat at both ends of the semiconductor plate and in the semiconductor pieces, which have higher resistances than the metal sample. While a complex current having a component along the semiconductor plate would also give a non-zero contribution to the heating, but this effect is negligible.

\begin{figure}[htbp]
    \centering
    \includegraphics[width=1.0\linewidth]{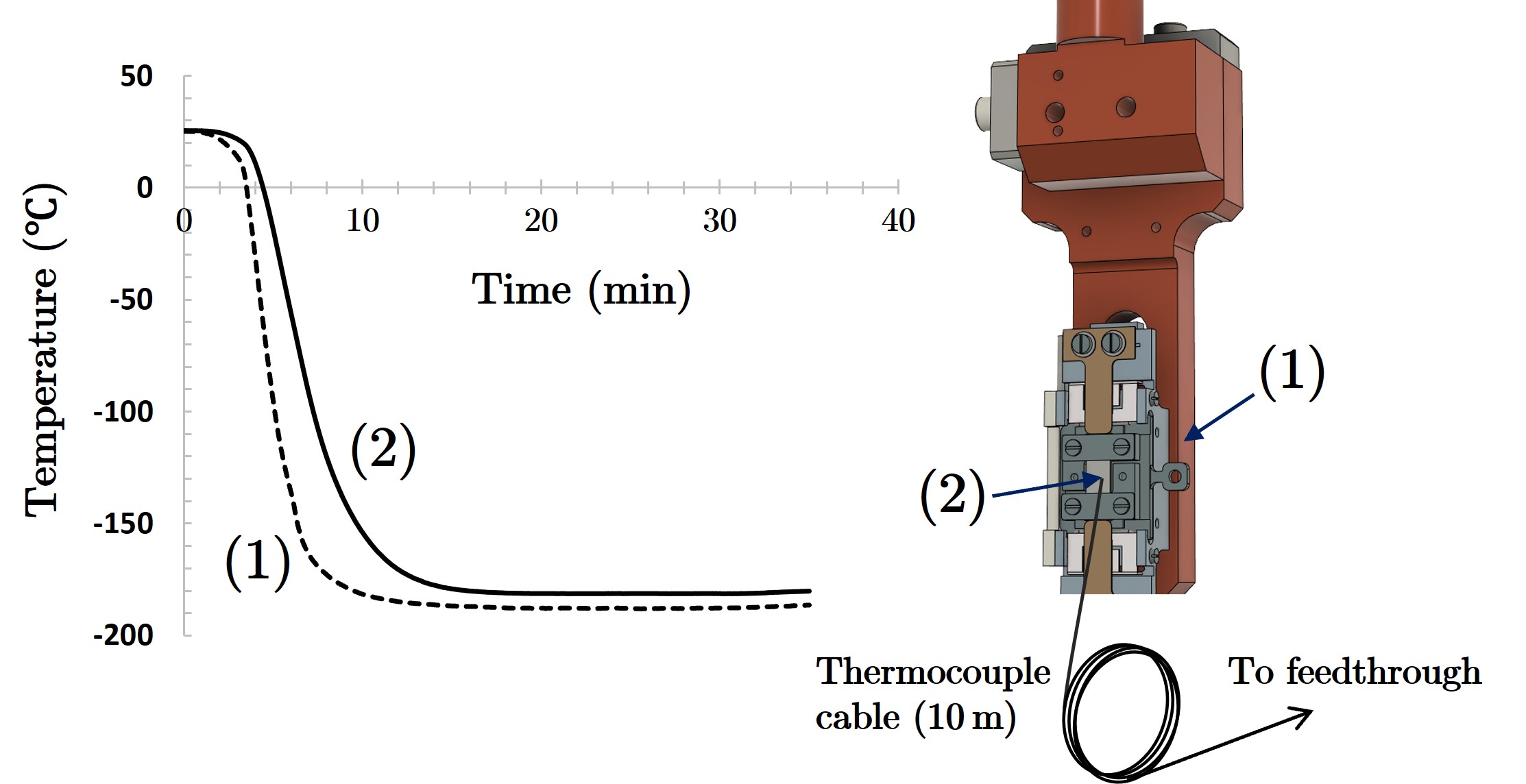}
    \caption{(Left) Temperature variation during cooling with liquid N$_2$ at two positions on the manipulator in the LEPD measurement chamber: (1) near the sample-plate receiver and (2) at the surface of a dummy silicon sample. (Right) Schematic diagram showing the positions of the thermocouples used for these measurements. The surface temperature was measured using a \diameter\qty{0.1}{mm} type-K thermocouple with a \qty{10}{m} long wire, which was looped inside the chamber to minimize heat conduction. The wire was then wrapped with several layers of super-insulating sheets (not shown in the figure) to reduce radiant heat. }
    \label{fig:samplecooling}
\end{figure}

To estimate the sample temperature during the experiments, a permanently installed type-K thermocouple was attached near the sample plate receiver of the manipulator in the LEPD measurement chamber, as indicated by position (1) on the right side of Fig.~\ref{fig:samplecooling}. We calibrated the surface temperature using a \diameter\qty{0.1}{mm} type-K thermocouple directly affixed to a dummy sample with aluminum adhesive tape at position (2) on the right side of Fig.~\ref{fig:samplecooling}. To minimize heat conduction from the feedthrough, a \qty{10}{m} long thermocouple wire was used. It was looped inside the chamber and wrapped in a bag made of several layers of super-insulating sheet to reduce radiant heat. The left side of Fig.~\ref{fig:samplecooling} shows the temperature change as the sample was cooled from room temperature using liquid N$_2$. The sample surface on the LEPD manipulator reached a stable temperature of \qty{-180}{\degreeCelsius} within \qty{15}{minutes}.

\section{Experimental}

\subsection{Evaluation of the LEPD experimental station}
\label{subsection:samplepreparation}

\begin{figure}[htbp]
    \centering
    \includegraphics[width=1.0\linewidth]{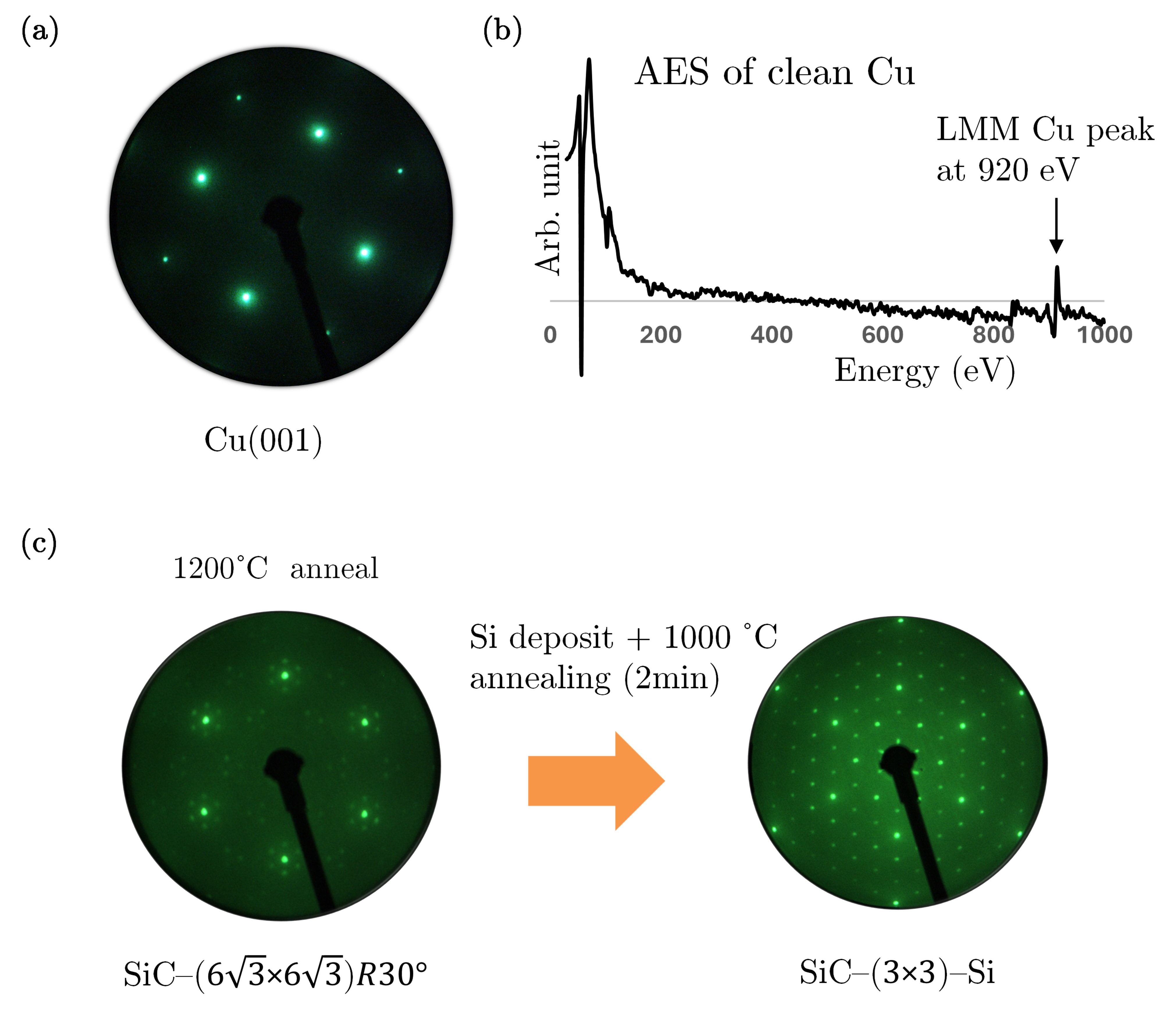}
    \caption{(a) The LEED pattern of Cu(001)--$(1\!\times\!1)$ at \qty{140}{eV} after cleaning with Ar$^+$ sputtering (voltage \qty{800}{\eV}, current \qty{30}{\micro\ampere} for \qty{15}{minutes} at Ar pressure $\qty{1.3e-3}{\Pa}$) and annealing to \qty{600}{\degreeCelsius} for \qty{15}{mins}. (b) An AES spectrum confirming the cleanliness of the Cu(001) sample, showing a sharp LLM (Cu Auger transition) peak at a binding energy of \qty{920}{eV}. (c) The diffraction pattern of the SiC--6H--(0001)--$(6\sqrt{3}\!\times\!6\sqrt{3})R\qty{30}{\degree}$ after flashing at \qty{1200}{\degreeCelsius}, and the $(3\!\times\!3)$ diffraction pattern after Si deposition at \qty{1000}{\degreeCelsius} for \qty{2}{min}.}
    \label{fig:leedpatterns}
\end{figure}

A Cu(001) surface was cleaned in the sample preparation chamber using Ar$^+$ sputtering (voltage \qty{800}{\eV}, current \qty{35}{\micro\ampere} for \qty{15}{minutes} at Ar pressure $\qty{1.3e-3}{\Pa}$), followed by resistive heating to \qty{600}{\degreeCelsius} for \qty{15}{minutes}. The sample was fixed with a silicon carbide (SiC) plate underneath and two small SiC pieces on the top surface at both ends of the sample, corresponding to the ``semiconductor plate" and ``semiconductor pieces" described previously and shown in Fig.~\ref{fig:sampleholder}~(d) and Fig.~\ref{fig:sampleholder}~(e). After this cleaning, a sharp $1\!\times\!1$ LEED pattern was observed over a broad energy range; a typical pattern at \qty{140}{\eV} is shown in Fig.~\ref{fig:leedpatterns}~(a). The surface cleanliness was also confirmed by the AES spectrum, which shows a sharp LMM (Cu Auger transition) peak at a binding energy of \qty{920}{\eV}, as presented in Fig.~\ref{fig:leedpatterns}~(b). The AES spectrum showed no Si LVV peaks, indicating the absence of Si contamination after the heating process at \qty{600}{\degreeCelsius}, which is well below the desorption threshold of Si from SiC (${\sim}\qty{1030}{\degreeCelsius}$) ~\cite{forbeaux1999high}.

Additionally, a 6H--SiC(0001) sample was used to demonstrate both the heating capabilities and deposition functionality of the electron beam evaporator. Fig.~\ref{fig:leedpatterns}~(c) shows the 6H--SiC--(0001)--$(6\sqrt{3}\!\times\!6\sqrt{3})R\ang{30}$ LEED pattern after flashing the sample to \qty{1200}{\degreeCelsius}, and the SiC--$(3\!\times\!3)$ structure obtained following Si deposition at \qty{1000}{\degreeCelsius}. A thermal pyrometer was used to estimate the sample surface temperature during the heat treatment through a viewing port from the outside of the chamber. 

\begin{figure}[htbp]
    \centering
    \includegraphics[width=1.0\linewidth]{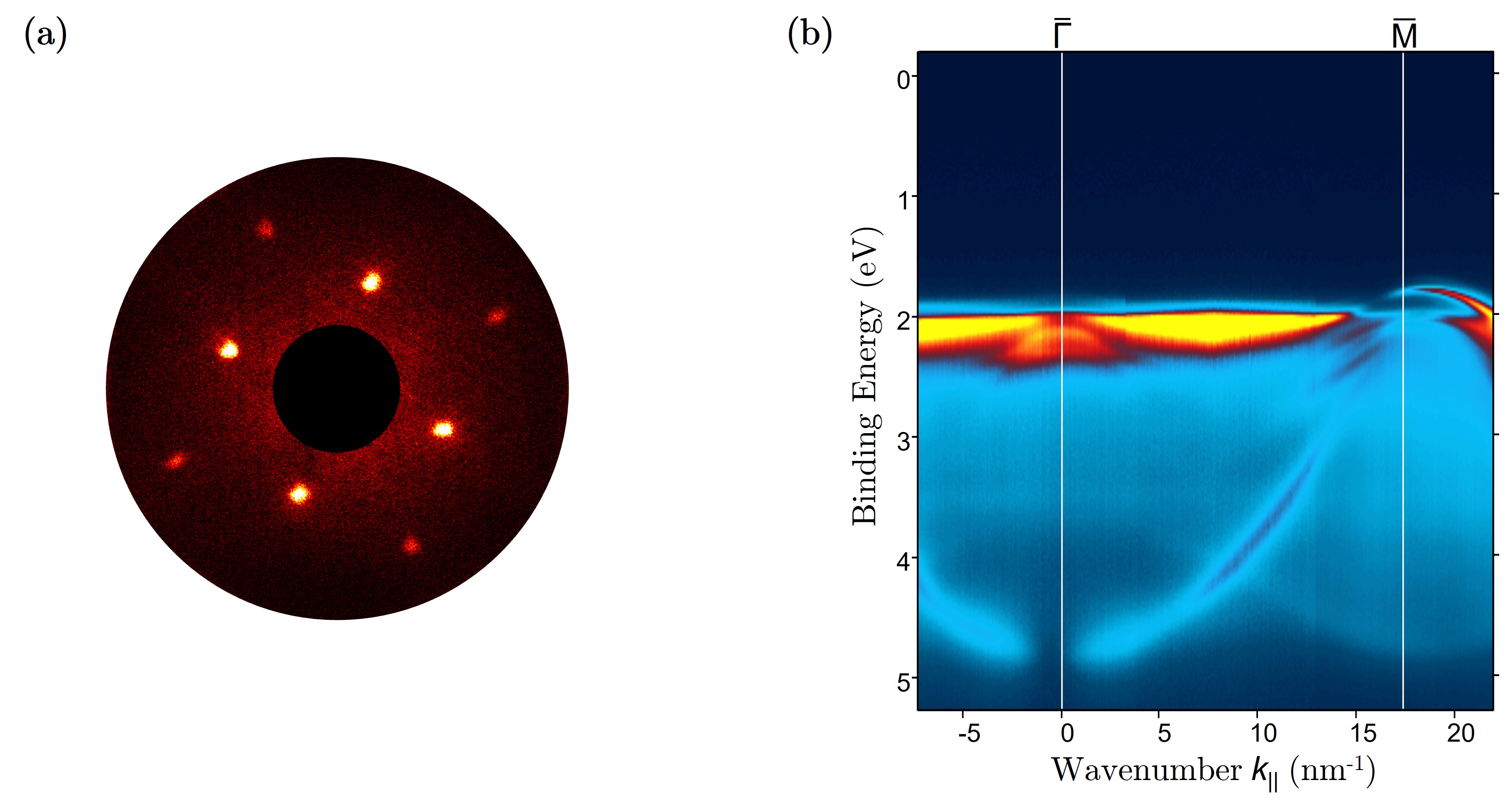}
    \caption{(a) LEPD pattern of Cu(001) at \qty{171}{eV}. The experiment was carried out at SPF-A4 of SPF, IMSS, KEK. (b) An ARPES map of the Cu(001) sample indicating the binding energy as a function of the wavenumber $k_\parallel$ along the $\overline{\Gamma}$--$\overline{\text{M}}$ direction.  The experiment was carried out with a sample prepared at SPF-A4 and transferred to BL-13B of PF, IMSS, KEK using a portable UHV transfer vessel.}
    \label{fig:lepdandarpes}
\end{figure}

An example of a LEPD pattern decoded from the data obtained by the Hexanode for Cu(001) is shown in Fig.~\ref{fig:lepdandarpes}~(a). The data acquisition time for each diffraction pattern was \qty{60}{seconds}, which is two orders of magnitude faster than previously reported~\cite{wada2018observation}. This improvement is attributed to a tenfold increase in the LEPD beam intensity, achieved through the optimization of beam generation and tuning, as well as improved quality of the sample surface. Due to the detector configuration, consisting of a flat MCP and a Hexanode, the appearance of the spots in Fig.~\ref{fig:lepdandarpes}~(a) is slightly elongated in the radial direction compared to those from a conventional LEED system with a hemispherical screen. However, this elongation does not affect the $I$-$V$ analysis, as it tracks the intensity variations of selected diffraction spots over different energies. LEPD $I$-$V$ data for Cu(001) have been extracted from the diffraction patterns. The extracted $I$-$V$ curves and subsequent structure analysis will be reported elsewhere.

\subsection{ARPES observation}

After the LEPD observation, the Cu(001) surface was re-cleaned in the LEPD preparation chamber using the same well-established and reproducible procedure described in section~\ref{subsection:samplepreparation}. Following the surface cleaning, the sample was transferred through the load-lock chamber into a portable UHV transfer vessel, capable of carrying up to three samples simultaneously~\cite{kitamura2024}. The sample was then transported to station BL-13B at the PF (IMSS, KEK) on the same KEK Tsukuba campus, where ARPES observations were promptly conducted. A clear band structure of the clean Cu(001) surface was observed. An ARPES intensity plot of the Cu(001) sample along the $\overline{\Gamma}$--$\overline{\text{M}}$ direction is shown in Fig.\ref{fig:lepdandarpes}~(b). The data were obtained using s-polarized light with a photon energy of \qty{100}{eV} at an incidence angle of \ang{65}, at room temperature. The wavenumber $k_\parallel = \qty{0}{nm^{-1}}$ corresponds to the center of the surface Brillouin zone, and the $\overline{\text{M}}$ point is at $k_\parallel = \qty{17.4}{nm^{-1}}$. The results are consistent with previous studies~\cite{roth2016angle}.

When consecutive measurements on a sample are scheduled, as in the case detailed here---LEPD at SPF-A4 and ARPES at PF BL-13B---the sample holder assembly is slightly modified from the standard design. One of the sapphire sheets in the original design is replaced with a Mo sheet of exactly the same shape and size. This modification is necessary to meet the electrical grounding requirements for ARPES experiments at PF BL-13B and to realize the seamless execution of both LEPD and ARPES measurements on the same sample.

\section{Summary}

We have successfully upgraded the LEPD experimental station at SPF, IMSS, KEK, to enhance its capabilities for surface structure determination. The data acquisition time for each diffraction pattern of Cu(001) was reduced to one minute, enabling the collection of multiple patterns required for LEPD $I$-$V$ measurements within a couple of hours, thus making high-quality measurements feasible before surface degradation occurs. The upgraded station consists of two dedicated UHV chambers: a sample preparation chamber and a LEPD observation chamber, with the latter connected to the SPF-A4 beamline. The sample preparation chamber is equipped with an Ar$^+$ sputtering system, a triple-pocket electron beam evaporator, three gas introduction systems, with additional ports available if needed, as well as a LEED/AES system. 

This system enables both LEPD and ARPES measurements on the same sample under consistent surface conditions, achieved by using identical preparation environments and procedures. 

\section*{CRediT authorship contribution statement}

R. Ahmed: Formal analysis, Investigation, Methodology, Resources, Visualization, Writing - original draft. I. Mochizuki: Methodology, Investigation, Resources, T. Hyodo: Conceptualization, Writing - review \& editing. T. Shirasawa: Methodology, Resources, Software, Writing - review \& editing. S. Mizuno: Methodology, Resources, Software. Y. Kondo: Investigation, Methodology, Resources. K. Ozawa: Investigation, Methodology, Resources, Writing - review \& editing. M. Kitamura: Methodology, Resources. K. Amemiya: Resources. B. Checinski: Resources. J. Ociepa: Resources. A. Czasch: Formal analysis, Resources, Software. O. Jagutzki: Formal analysis, Resources, Software. K. Wada: Conceptualization, Data curation, Formal analysis, Funding acquisition, Investigation, Methodology, Project administration, Resources, Software, Writing - original draft.

\section*{Declaration of competing interest}

The authors declare that they have no known competing financial interests or personal relationships that could have appeared to influence the work reported in this paper.

\section*{Acknowledgment}

This work was supported by JSPS KAKENHI Grant Numbers JP23K21832, JP21H03745, and JP18H03476. Experiments were conducted under the approval of the PF PAC (Proposal Nos.~2023G656, 2023Q003, 2022Q003, 2021G681, 2019G692, 2016S2-006). The authors also gratefully acknowledge the KEK Mechanical Engineering Center for their support in hydrogen furnace processing of Ni thin films for positron remoderation. We thank Dr.\ N.\ Zafar for discussions.

% \appendix
% \section{Sample Appendix Section}
% \label{sec:sample:appendix}
% Lorem ipsum dolor sit amet, consectetur adipiscing elit, sed do eiusmod tempor section \ref{sec:sample1} incididunt ut labore et dolore magna aliqua. Ut enim ad minim veniam, quis nostrud exercitation ullamco laboris nisi ut aliquip ex ea commodo consequat. Duis aute irure dolor in reprehenderit in voluptate velit esse cillum dolore eu fugiat nulla pariatur. Excepteur sint occaecat cupidatat non proident, sunt in culpa qui officia deserunt mollit anim id est laborum.

\bibliographystyle{elsarticle-num} 
\biboptions{sort&compress}
\bibliography{cas-refs}
% \input{rezwan2025development.bbl}





% \end{thebibliography}
\end{document}